\documentstyle{mn}

\newif\ifAMStwofonts

\ifoldfss
  \ifCUPmtlplainloaded \else
    \NewTextAlphabet{textbfit} {cmbxti10} {}
    \NewTextAlphabet{textbfss} {cmssbx10} {}
    \NewMathAlphabet{mathbfit} {cmbxti10} {} 
    \NewMathAlphabet{mathbfss} {cmssbx10} {} 
  \fi
  \ifAMStwofonts
    \ifCUPmtlplainloaded \else
      \NewSymbolFont{upmath} {eurm10}
      \NewSymbolFont{AMSa} {msam10}
      \NewMathSymbol{\upi}     {0}{upmath}{19}
      \NewMathSymbol{\umu}     {0}{upmath}{16}
      \NewMathSymbol{\upartial}{0}{upmath}{40}
      \NewMathSymbol{\leqslant}{3}{AMSa}{36}
      \NewMathSymbol{\geqslant}{3}{AMSa}{3E}

      \let\leq=\leqslant 
       
    \fi
  \fi
\fi 

\ifnfssone
  \newmathalphabet{\mathit}
  \addtoversion{normal}{\mathit}{cmr}{m}{it}
  \addtoversion{bold}{\mathit}{cmr}{bx}{it}
  \newmathalphabet{\mathbfit} 
  \addtoversion{normal}{\mathbfit}{cmr}{bx}{it}
  \addtoversion{bold}{\mathbfit}{cmr}{bx}{it}
  \newmathalphabet{\mathbfss} 
  \addtoversion{normal}{\mathbfss}{cmss}{bx}{n}
  \addtoversion{bold}{\mathbfss}{cmss}{bx}{n}
  \ifAMStwofonts
    \ifCUPmtlplainloaded \else
      \UseAMStwoboldmath
      \makeatletter
      \new@mathgroup\upmath@group
      \define@mathgroup\mv@normal\upmath@group{eur}{m}{n}
      \define@mathgroup\mv@bold\upmath@group{eur}{b}{n}
      \edef\UPM{\hexnumber\upmath@group}
      \new@mathgroup\amsa@group
      \define@mathgroup\mv@normal\amsa@group{msa}{m}{n}
      \define@mathgroup\mv@bold\amsa@group{msa}{m}{n}
      \edef\AMSa{\hexnumber\amsa@group}
      \makeatother
      \mathchardef\upi="0\UPM19
      \mathchardef\umu="0\UPM16
      \mathchardef\upartial="0\UPM40
      \mathchardef\leqslant="3\AMSa36
      \mathchardef\geqslant="3\AMSa3E

      \let\leq=\leqslant 

    \fi
  \fi
\fi 

\ifnfsstwo
  \DeclareMathAlphabet{\mathbfit}{OT1}{cmr}{bx}{it}
  \SetMathAlphabet\mathbfit{bold}{OT1}{cmr}{bx}{it}
  \DeclareMathAlphabet{\mathbfss}{OT1}{cmss}{bx}{n}
  \SetMathAlphabet\mathbfss{bold}{OT1}{cmss}{bx}{n}
  \ifAMStwofonts
    \ifCUPmtlplainloaded \else
      \DeclareSymbolFont{UPM}{U}{eur}{m}{n}
      \SetSymbolFont{UPM}{bold}{U}{eur}{b}{n}
      \DeclareSymbolFont{AMSa}{U}{msa}{m}{n}
      \DeclareMathSymbol{\upi}{0}{UPM}{"19}
      \DeclareMathSymbol{\umu}{0}{UPM}{"16}
      \DeclareMathSymbol{\upartial}{0}{UPM}{"40}
      \DeclareMathSymbol{\leqslant}{3}{AMSa}{"36}
      \DeclareMathSymbol{\geqslant}{3}{AMSa}{"3E}

      \let\leq=\leqslant 

    \fi
  \fi
\fi 

\ifCUPmtlplainloaded \else
  \ifAMStwofonts \else 
    \def\upi{\pi}
    \def\umu{\mu}
    \def\upartial{\partial}
  \fi
\fi

\def\approxlt{\mathrel{\hbox{\rlap{\lower.55ex \hbox {$\sim$}}
        \kern-.3em \raise.4ex \hbox{$<$}}}}

\title[Peculiar, Low Luminosity Type II Supernovae]
{Peculiar, Low Luminosity Type II Supernovae:\\
Low Energy Explosions in Massive Progenitors?
\thanks{Based on observations collected at the European Southern Observatory
and Cerro Totolo Inter-American Observatory, Chile (Proposals ESO 63.H-0141
and 66.D-0683).}}

\author[L. Zampieri et al.]
       {L. Zampieri$^{1}$, A. Pastorello$^{1,2,3}$, M. Turatto$^{1}$,
       E. Cappellaro$^{4}$, S. Benetti$^{1}$,\and
       G. Altavilla$^{1,2}$, P. Mazzali$^{5}$, M. Hamuy$^{6}$
       \\
$^1$INAF - Astronomical Observatory of Padova, Vicolo dell'Osservatorio 5,
I-35122 Padova, Italy \\
$^2$Department of Astronomy, University of Padova, Vicolo dell'Osservatorio 2,
I-35122 Padova, Italy\\
$^3$Department of Physics and Astronomy, University of Oklahoma,
Norman, OK 73019, USA\\
$^4$INAF - Astronomical Observatory of Capodimonte, Via Moiariello 16,
I-80131 Napoli, Italy\\
$^5$INAF - Astronomical Observatory of Trieste, Via Tiepolo 11,
I-34131 Trieste, Italy\\
$^6$ The Observatories of the Carnegie Institution of
Washington, 813 Santa Barbara Street, Pasadena, CA 91101, USA}

\date{Accepted ...
      Received ...;
      in original form ...}

\pagerange{\pageref{firstpage}--\pageref{lastpage}}
\pubyear{2001}

\begin{document}

\maketitle

\label{firstpage}

\begin{abstract}
A number of supernovae, classified as Type II, show remarkably
peculiar properties such as an extremely low expansion velocity and an
extraordinarily small amount of $^{56}$Ni in the ejecta. We present a
joint analysis of the available observations for two of these
peculiar Type II supernovae, SN 1997D and SN 1999br, using a
comprehensive semi-analytic method that can reproduce the light curve
and the evolution of the line velocity and continuum temperature. We
find that these events are under-energetic with respect to a typical
Type II supernova and that the inferred mass of the ejecta is
relatively large. We discuss the possibility that these supernovae
originate from the explosion of a massive progenitor in which the rate
of early infall of stellar material on the collapsed core is
large. Events of this type could form a black hole remnant, giving
rise to significant fallback and late-time accretion.
\end{abstract}

\begin{keywords}
supernovae: general -- supernovae: individual: SN 1997D --
supernovae: individual: SN 1999br -- methods: analytical

\end{keywords}

\section{Introduction}


Recently, a number of supernovae have been discovered that, according
to their spectral properties, are classified as Type II but that, at
the same time, show remarkably peculiar properties. The first clearly
identified example was the exceptionally faint SN 1997D in NGC 1536
(Turatto et al. \cite{tetal98}; Benetti et al. \cite{betal01}).  SN
1997D showed both a very faint radioactive tail in the light curve,
indicating an ejected $^{56}$Ni mass of only a few $\times 10^{-3}
M_\odot$, and a low expansion velocity of $\sim 1000$ km
s$^{-1}$. Modeling the spectra and light curve Turatto et
al. \cite{tetal98} concluded that a low energy explosion in a 26
$M_\odot$ progenitor star could successfully fit the early
observations. The scenario of the low energy explosion of a high mass
progenitor is also consistent with the late time ($\sim 400$ days)
spectral and photometric data (Benetti et al. \cite{betal01}).  Chugai
\& Utrobin \cite{cu00} have presented an alternative analysis in which
the progenitor was a star at the low end of the mass range of
core-collapse supernovae (8--12 $M_\odot$).

The very low luminosity at discovery (10 times less than the peak
luminosity reached by SN 1987A) and very small expansion velocity of
the ejecta (3-4 times less than that of a normal Type II) of SN 1997D
suggest that this explosion event was under-energetic. The mechanism
that causes the energy of the explosion to be so low is a challenging
and important problem in the physics of core-collapse supernovae. In
fact, the exact evolution of the star after neutrino reheating depends
on the rate of early infall of stellar material on the collapsed core
and on the binding energy of the envelope. If both are large, as in
high mass stars ($M>20 M_\odot$), the energy available to accelerate
and heat up the ejecta can be greatly reduced and, eventually, may
become insufficient to cause a successful explosion. Therefore, after
the passage of the shock wave (and possibly the reverse shock formed
at the H--He interface) a variable amount of matter may remain
gravitationally bound to the collapsed remnant and fall back onto it
(Woosley \& Weaver \cite{ww95}). For this reason it was early
suggested that SN 1997D could host a black hole remnant formed during
the explosion (Turatto et al. \cite{tetal98}) and that the luminosity
powered by fallback of envelope material onto the central black hole
could emerge at about 1000-1200 days after the explosion (Zampieri,
Shapiro \& Colpi \cite{zsc98}). Unfortunately, to date it has not been
possible to confirm or disprove observationally this prediction.

After SN 1997D, a number of supernovae with similar observational
properties have been identified. These objects appear to define a
fairly homogeneous group of explosion events (Pastorello et
al. \cite{petal02}). As for SN 1997D, these supernovae provide a
unique opportunity to probe the physics of the explosion and reach a
better understanding of both the explosion mechanism and the
conditions for the formation of black hole remnants. In this Paper we
present the results of a joint analysis of the observations of two
peculiar Type II supernovae with very low luminosity and expansion
velocity, SN 1997D and SN 1999br, which are representative of the
properties of the whole group. In particular, SN 1999br represents the
most extreme case of a low-luminosity event to date and appears to
follow a recently reported correlation between expansion velocities of
the ejecta and bolometric luminosities during the plateau phase (Hamuy
\& Pinto \cite{hp02}).

In Section~\ref{sec1} we describe the basic spectral and photometric
data of these supernovae. Section~\ref{sec2} summarizes the
semi-analytic method employed to model the light curve and the
evolution of the line velocity and continuum temperature of Type II
supernovae. Finally, in section~\ref{sec3} the main results are
presented and their consequences regarding the nature of the
progenitors and the energy of the explosion are briefly discussed.

\section{Peculiar Type II supernovae}
\label{sec1}

\begin{figure}
 \vspace{8.0truecm}
 \includegraphics{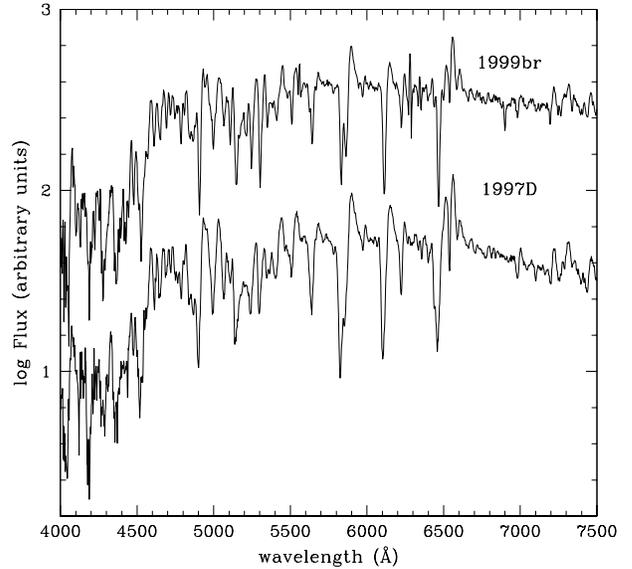}
 \caption{Spectra of SN 1997D and SN 1999br at a phase $\sim 100$ days.  The
spectrum of SN 1999br was obtained on 1999 July 20 at the ESO-Danish
1.54m telescope equipped with DFOSC (+grism 4). The spectrum of SN
1997D was taken on 1997 January 15 with the ESO 1.52m telescope
(B\&C+grating 27).
}
 \label{fig1}
\end{figure}

Supernova 1997D, serendipitously discovered on 14 January 1997 during
an observation of the parent galaxy NGC 1536 (De Mello \& Benetti
\cite{db97}), is the first, clearly identified example of a peculiar
Type II supernova with a very low luminosity and expansion velocity
(Turatto et al. \cite{tetal98}). It was detected when it was already
decaying from the plateau and was at least 2 mag fainter than a
typical Type II supernova (Patat et al. \cite{patetal94}). The decline
rate of the last segment of the bolometric light curve is consistent
with complete thermalization of the gamma rays from the radioactive
decay of $^{56}$Co into $^{56}$Fe. Assuming for SN 1997D the same
deposition as in SN 1987A, the ejected $^{56}$Ni mass is $\sim
10^{-3}-10^{-2} M_\odot$ (Benetti et al. \cite{betal01}).  The spectra
are dominated by a red continuum and P-Cygni profiles of H I, Ba II,
Ca II, Na I and Sc II (see Figure~\ref{fig1}).  The most striking
property of these spectra is the very low expansion velocity inferred
from the spectral lines.  The minima of the absorption lines give an
expansion velocity of about 1100-1200 km s$^{-1}$ for H (see Benetti
et al. \cite{betal01} for details).

Recently, other supernovae with properties similar to those of SN
1997D have been identified. A comprehensive analysis of the
observations of these peculiar supernovae, including the recently
discovered SN 2001dc, will be presented elsewhere (Pastorello et
al. \cite{petal02}). Here we focus on one particularly representative
object, SN 1999br in NGC 4900, discovered on 12 April 1999 (King
\cite{king99}). Filippenko et al. \cite{fetal99} pointed out
the exceptionally low luminosity of this event. Later, Patat et
al. \cite{patatetal99} recognized that SN 1999br shows other
properties similar to those of SN 1997D, such as spectra with narrow
P-Cygni lines, prominent Ba II features and a low continuum
temperature. This supernova had a long plateau lasting at least 110
days with a mean luminosity of only $\approxlt 10^{41}$ erg s$^{-1}$.
It was extensively monitored during the first $\sim 110$ days with the
CTIO and ESO telescopes, yielding extremely good photometric and
spectroscopic coverages (Hamuy et al. \cite{hametal02}). The spectrum
of SN 1999br taken at a phase of $\sim 100$ days shows narrow metal
lines (Figure~\ref{fig1}). As in the case of SN 1997D, the red
continuum and the strength of Ba II lines are caused by the low
temperature of the ejecta. The U-B bands are strongly affected by line
blanketing as the temperature decreases.

As shown in Figure~\ref{fig1}, the spectrum of SN 1997D obtained soon
after discovery and that of SN 1999br at $\sim 100$ days are
strikingly similar in both the continuum and the line
components. Taking the ratio of the two spectra results in a rather
flat curve where the main deviations are caused by slightly different
line width (i.e. different expansion velocities of the ejecta) and
spectral resolution. The expansion velocity of SN 1997D is only $\sim
5\%$ larger than that of SN 1999br. The close resemblance of these two
spectra, the low inferred expansion velocity ($\sim$1000 km s$^{-1}$)
and the fact that the two supernovae have comparable luminosities
strongly suggest that SN 1997D and SN 1999br may be similar events. In
the following we will investigate in detail the consequences of this
assumption.

SN 1997D was discovered at the end of the plateau stage when the light
curve was plummeting. The duration and luminosity of the plateau were
inferred by comparison between models and observations and the
consequent estimate of the explosion epoch ($\sim 60$ days before
discovery) was uncertain (Turatto et al. \cite{tetal98}). On the other
hand, the date of the explosion of SN 1999br has an uncertainty of
only a few days (thanks to a stringent pre-discovery limit). Then, from
the similarity of the observational properties (luminosities and
spectra) of SN 1997D and SN 1999br, we tentatively assume that the
phase of SN 1997D at discovery is $\sim 90-100$ days. This assumption
will be checked by means of a detailed comparison of
spectral-synthesis models with observations, presently under way.

\section{The Model}
\label{sec2}

In order to determine the physical properties of a supernova, the
observed light curve and spectra are compared to numerical
simulations. Usually, it is important to carry out a preliminary
investigation in order to obtain an approximate but reliable estimate
of the physical conditions of the ejected gas and to establish a
framework in which more detailed follow-up calculations can be
performed. For these reasons, we have implemented a semi-analytic
model in spherical symmetry that can provide a robust estimate of the
parameters of the ejecta of Type II supernovae. The novelty of the
present approach lies mainly in the fact that the physical properties
of the envelope are derived by performing a simultaneous comparison of
the observed and simulated light curve, evolution of the line velocity
and continuum temperature. The model follows the approach originally
introduced by Arnett \cite{arn82} and later developed by Arnett \& Fu
\cite{arfu89} and Arnett \cite{arn96}. The treatment of the motion of
the recombination front follows in part the work of Popov
\cite{pop92}. The present analysis includes all the relevant energy
sources powering the supernova and evolves the envelope in 3 distinct
phases that cover the whole evolution from the photospheric up to the
late nebular stages.
We include also the energy input from recombination that may be
particularly important for low energy events like SN 1997D. The model
has been tested against numerical radiation-hydrodynamic computations
in spherical symmetry under the same assumptions, giving good
agreement. The details of the method and its validation will be
presented in a companion paper (Zampieri et al. \cite{zetal02}). Here
we briefly summarize the basic ideas.

\begin{figure}
 \vspace{11.0truecm}
 \includegraphics{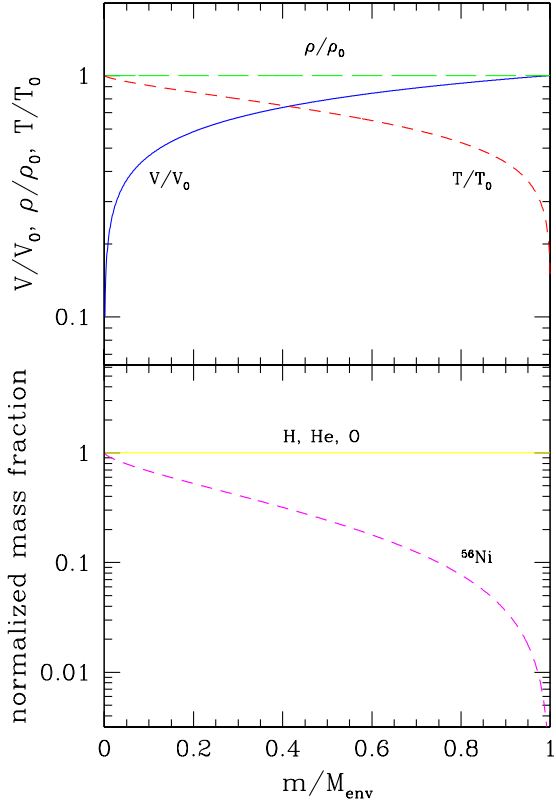}
 \caption{Initial ($t=t_{in}$) conditions of the post-shock envelope
in our model. {\it Upper panel}: velocity $V$, density
$\rho$ and temperature $T$ distributions as a function
of interior (Lagrangian) mass $m$. {\it Bottom panel}: Dependence of
Hydrogen, Helium, Oxygen and Nickel abundances on $m$.}
 \label{figdis}
\end{figure}

Full hydrodynamical calculations using realistic pre-supernova models
show that the dynamical evolution of the envelope during shock passage
is quite complex. The propagation of the shock determines how the
explosion energy is distributed in the envelope of the progenitor
star. The innermost layers comprised of helium and heavier elements
transfer almost all of their kinetic energy and momentum to the
Hydrogen envelope. The star mixes but does not homogenize. The actual
velocity, density and heavy element distributions of the post-shock
material affect the light curve and estimation of the envelope mass in
a major way. In the present analysis we do not consider the evolution
of the star during this complex phase but rather assume idealized
initial conditions that provide an approximate description of the
ejected material after shock (and possible reverse shock) passage, as
derived from hydrodynamical calculations. The evolution starts at time
$t_{in}$ (since core collapse) when the envelope is essentially
free-coasting and in homologous expansion. In realistic explosion
calculations, it takes several hours in order for the envelope to
relax to this state. We adopt $t_{in}\simeq 5$ hrs. At this stage, the
velocity $V$ of each gas shell is approximately constant and
proportional to its position $r$, $V(r)=V_0 (r/R_0)$, where $R_0$ and
$V_0$ are the initial radius and velocity of the outermost shell of
the envelope. Therefore, the radius of each gas shell increases
linearly with time: for the outermost shell, $R=R_0+V_0
(t-t_{in})$. The post-shock, ejected envelope is assumed to have
spatially constant density $\rho$ and total mass $M_{env}=4\pi \rho
R^3/3$. Mass conservation gives $\rho = \rho_0 (R_0/R)^3$, where
$\rho_0$ is the initial density. In reality, the outer part of the
star develops a steep power-law density structure that affects the
light curve during the first few days after shock breakout (which is
not included). However, this is only $\sim$ 1\% by mass of the star,
while most of the stellar material resides in the inner part with
roughly constant density. The latter region dominates the light curve
after 10--20 days. The expansion velocity of the envelope as a
function of interior (Lagrangian) mass $m(r)=4\pi \rho r^3/3$ is then
$V(m)=V_0 (m/M_{env})^{1/3}$ (see Figure~\ref{figdis}). The initial
thermal+kinetic energy of the envelope is $E=(3/10)M_{env}V_0^2/f_0$, where
$f_0$ is the initial fraction of kinetic energy. Elements are assumed
to be completely mixed throughout the envelope.  In our simple
spherically symmetric model, their distribution depends only on $r$
(or $m$). In particular, Hydrogen, Helium and Oxygen are assumed to be
uniformly distributed, whereas $^{56}$Ni is more centrally peaked (see
Figure~\ref{figdis}). Our assumptions about the velocity and elemental
distributions provide only an approximate description of the actual
post-shock structure of the envelope and should be regarded as a
potential source of systematic uncertainty in the present model.

The evolution of the supernova envelope starts at time $t_{in}$ and is
schematically divided into 3 phases. During the first stage (a few
tens of days) the envelope is hot and completely ionized because of
the energy deposited by the shock wave. In the following phase (from
30--40 days up to $\sim 100$ days), because of the decrease in
temperature caused by expansion and radiative diffusion, the envelope
recombines and can be schematically divided into two regions, below
and above the position of the recombination wavefront. In the third
stage (after $\sim 100$ days), the ejecta are completely recombined
and transparent to optical photons. The evolution is computed solving
the energy balance equation for the envelope gas. The thermal balance
is governed by the competition among the energy input from trapped
gamma-rays, the $PdV$ work and the energy losses through radiative
diffusion. Assuming that radiation is in LTE with the gas throughout
the envelope, the energy balance equation becomes a second order
partial differential equation for the temperature $T$. Arnett
\cite{arn96} has shown that, under specific assumptions on the
spatial distribution of $^{56}$Ni, the energy equation (with
appropriate boundary conditions) can be solved by separation of
variables. The function that describes the radial dependence $\psi(r)$
is solution of an eigenvalue equation (see e.g. equation~[D.48] of
Arnett \cite{arn96}). The final solution has the form
\begin{equation}
T^4(t,r) = T_0^4 \left(\frac{R_0}{R}\right)^4 \psi(x) \phi(t) \, ,
\label{temp}
\end{equation}
where $x=r/R$, $T_0$ is a constant reference temperature, $\phi(t)$ is
the function that describes the time dependence [$\phi(0)=1$] and, in
the limit of zero mean free path, $\psi(x) = \sin(\pi x)/(\pi x)$
(Arnett's ``radiative zero'' solution). The term $(R_0/R)^4$ accounts
for adiabatic expansion. At $t=t_{in}$, the radial temperature profile
is $T/T_0=\psi(x)^{1/4}$ (see Figure~\ref{figdis}). Once a solution
for $\phi(t)$ is calculated, the emitted luminosity is approximately
given by the expression for the diffusion luminosity $L=- (4\pi r^2
c)/(3 \kappa \rho) [\partial a T^4/\partial r]_{r=R}$, where $\kappa$
is the gas opacity.

When the gas starts to recombine at $t=t_{rec,0}$, the part of the
envelope below the position $r_i$ of the wavefront is assumed to be in
LTE with radiation. We adopt the approximation that in this region
radiative diffusion can effectively readjust the radial temperature
distribution to the changing position of the outer boundary at
$r_i=r_i(t)$ (``slow approximation'', Arnett \& Fu \cite{arfu89}). In
this assumption, the spatial dependence of $T$ is essentially given by
Arnett's ``radiative zero'' solution. Therefore, we search for
approximate solutions of the energy equation of the form given in
equation~(\ref{temp}), with $x$ replaced by $x/x_i$, $x_i=r_i/R$ and
$\psi(x/x_i) = \sin(\pi x/x_i)/(\pi x/x_i)$ (Arnett \cite{arn96}).
The function that gives the time dependence of the temperature
profile, $\phi(t)$, can be approximately computed setting the
diffusion luminosity at recombination equal to the luminosity emitted
by a blackbody at the effective temperature, $4\pi r_i^2 \sigma
T^4_{eff}$ (Popov \cite{pop92}). This gives
\begin{equation}
\phi(t) = \frac{3}{4} \kappa \rho_0 R_0 \left( \frac{T_{eff}}{T_0} 
\right)^4
\left( - y^2 \frac{d\psi}{dy} \right)_{y=1}^{-1} 
\left(\frac{R}{R_0}\right)^2
x_i \, ,
\label{phixi}
\end{equation}
where $y=x/x_i$ and $T_{eff}$ is the effective temperature,
approximately constant and nearly equal to the gas ionization
temperature. Using equations~(\ref{temp}) (with $x$ replaced by
$x/x_i$) and~(\ref{phixi}), and integrating the energy equation over
the ionized region, we obtain an equation for the motion of the
recombination wavefront $x_i(t)$. Once a solution for $x_i$ is
computed, the bolometric luminosity from the inner envelope is
approximately given by (see Zampieri et al. \cite{zetal02} for
details)
\begin{equation}
L_{r_i} = L + 4\pi r_i^2 v_i (a T_{eff}^4/2 + \rho Q_{ion}) \, ,
\label{lumtot}
\end{equation}
where $L=- (4\pi c a T_0^4)/(3 \kappa \rho_0) [y^2 d\psi/dy]_{y=1} R_0
x_i \phi(t)$ is the diffusion luminosity, the second term on the right
hand side is the total advection luminosity released because of the
wavefront motion, $v_i={\dot x_i}R$ is the wavefront velocity relative
to the envelope gas and $Q_{ion}$ is the recombination energy per unit
mass.

In the recombined region ($r_i \leq r \leq R$), the deposition of
gamma ray photon energy through Comptonization and photoelectric
absorption of heavy elements is the dominant thermal and radiative
process. The optical luminosity emitted in this region originates by
the reprocessing of the gamma rays and becomes important only at the
end of the recombination stage (that coincides with the end of the
plateau), when the radioactive decay time of $^{56}$Co becomes
comparable to the expansion timescale. An approximate expression for
this luminosity can be obtained neglecting the internal energy and the
$PdV$ work in the energy equation and integrating it over the interval
$r_i \leq r \leq R$. The total bolometric luminosity of the envelope
during this phase is then
\begin{equation} L_{tot} = L_{r_i} +
4\pi \rho_0 R_0^3 X_{Ni,0} f(t) \int_{x_i}^1 x^2 \xi_{Ni}(x) dx \, .
\label{lumtot2}
\end{equation}
where $X_{Ni,0}$ and $\xi_{Ni}(x)$ are the central mass fraction and
radial distribution of $^{56}$Ni, $f(t) = [ 3.9\times 10^{10}
e^{-t/\tau_{Ni}} + 7.2 \times 10^9 (e^{-t/\tau_{Co}} -
e^{-t/\tau_{Ni}}) ]$ erg g$^{-1}$ s$^{-1}$, $\tau_{Ni} = 8.8$ days and
$\tau_{Co} = 111$ days are the Nickel and Cobalt decay times.
Following Arnett \cite{arn96}, we take $\xi_{Ni}(x)=\psi(x)=\sin(\pi
x)/(\pi x)$ (see Figure~\ref{figdis}, where the $^{56}$Ni abundance is
shown as a function of $m$).

In the third stage, when the envelope has completely recombined, the
$^{56}$Co radioactive decay time is shorter than the expansion
timescale and the luminosity is given by the second term in
equation~(\ref{lumtot2}) with $x_i=0$.

\begin{figure}
 \vspace{14.0truecm}
 \includegraphics{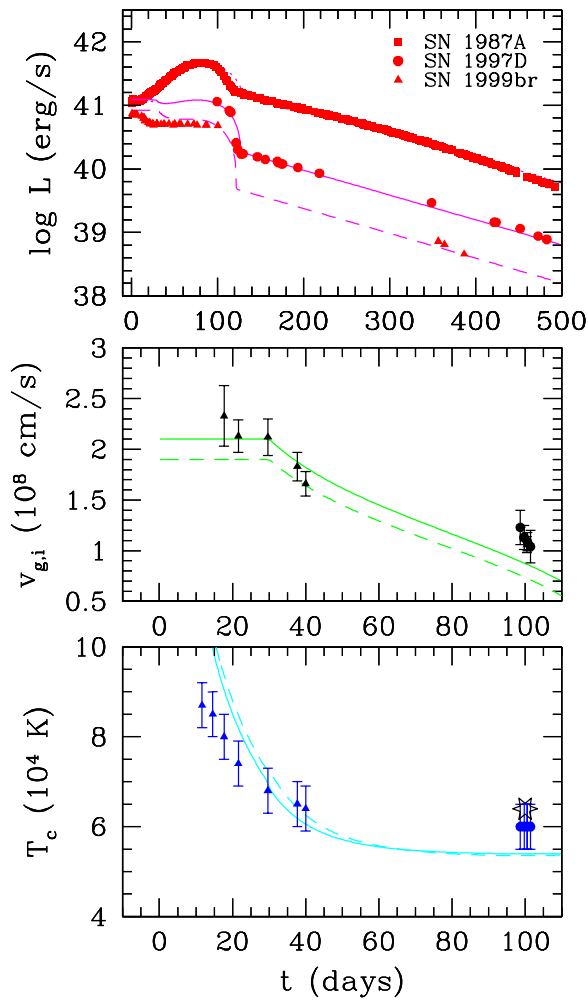}
 \caption{UBVRI luminosity, $L$, velocity of metal (Sc II) lines, $v_{g,i}$,
and continuum temperature, $T_c$, as a function of time for SN 1997D (circles)
and SN 1999br (triangles). The light curve of SN 1987A is also shown for
comparison. The adopted distance modulus and the estimated total reddening
are $\mu=18.49$, $E(B-V)=0.18$ for SN 1987A, $\mu=31.29$, $E(B-V)=0.02$
for SN 1997D and $\mu=31.19$, $E(B-V)=0.02$ for SN 1999br (see Pastorello
et al. (2002) for details).
The solid, long-dashed and short-dashed lines represent the
model curves. The velocities in the mid panel are those of the gas at the
position of the recombination wavefront $v_{g,i}=x_i V_0$. The asterisk is the
continuum temperature inferred from the spectral synthesis model of
SN 1997D at discovery (Turatto et al. (1998)) moved to a
phase of $\sim 100$ days.}
 \label{fig3}
\end{figure}

\section{Results and Discussion}
\label{sec3}

Following the discussion in \S~2, we assume that the phase of SN 1997D
at discovery is $\sim 90-100$ days and that the early light curve and
the evolution of the line velocity resemble closely those of SN
1999br. Figure~\ref{fig3} shows the evolution of the UBVRI luminosity,
the velocity of metal (Sc II) lines and the continuum temperature of
SN 1997D and SN 1999br along with the results of our semi-analytic
model calculations. In the top panel, the observed and calculated
light curves of SN 1987A are also shown for comparison. The continuum
temperatures have been estimated fitting the spectra with a Planckian
and are affected by a significant uncertainty ($\pm 500$ K) because of
the severe line blanketing at short wavelengths ($< 4500$ \AA, see
Figure~\ref{fig1}). The line velocity and continuum temperature
inferred from the model refer to the photospheric epoch.  Hence the
last two plots in Figure~\ref{fig3} are truncated at the end of the
recombination phase ($\sim 110$ days).

Considering the approximations adopted in the model, the general
agreement with observations is satisfactory. We emphasize that
obtaining a simultaneous ``fit'' of the light curve and the evolution
of the line velocity and continuum temperature is an essential
requirement to obtain a meaningful estimate of the parameters of the
ejected envelope. In particular, the long plateau, the apparent break
in the line velocity profile at $\sim 30$ days and its fast decline
impose rather severe constraints on the model. The break appears to be
related to the onset of recombination, that terminates when the light
curves plummets at $\sim 110$ days.

\begin{table*}
\caption{Parameters of the semi-analytic model}
\begin{flushleft}
\begin{tabular}{ccccccccccc}
\hline
 & $R_0$ & $M_{env}$ & $M_{Ni}$ & $V_0$ & $E$ &
 $f_0$ & $\kappa$ & $t_{rec,0}$ & $T_{eff}$ & $f_c$ \\
 & ($10^{12}$ cm) & ($M_\odot$) & ($M_\odot$) & ($10^8$ cm s$^{-1}$) &
 ($10^{51}$ erg) & & (cm$^2$ g$^{-1}$) & (days) & (K) \\
\hline
SN 1987A  & 5 & 18 & $7.5\times 10^{-2}$ & 2.7
          & 1.6 & 0.5 & 0.2 & 25 & 4800 & 1.1 \\
SN 1997D  & 9 & 17 & $8\times 10^{-3}$ & 2.1
          & 0.9 & 0.5 & 0.2 & 30 & 4400 & 1.2 \\
SN 1999br & 7.5 & 14 & $2\times 10^{-3}$   & 1.9
          & 0.6 & 0.5 & 0.2 & 30 & 4000 & 1.3 \\
\hline
\end{tabular}
\label{tab1}

$R_0$ is the initial radius of the envelope

$M_{env}$ is the ejected envelope mass

$M_{Ni}$ is the mass of $^{56}$Ni

$V_0$ is the velocity of the envelope material at the outer shell

$E$ is the initial thermal+kinetic energy of the ejecta

$f_0$ is the fraction of the initial energy that goes into kinetic
energy

$\kappa$ is the gas opacity

$t_{rec,0}$ is the time when the envelope starts to recombine

$T_{eff}$ is the effective temperature during recombination

$f_c=T/T_{eff}$ is the colour correction factor

\end{flushleft}
\end{table*}

The parameters of the post-shock, ejected envelope required to
reproduce the observations of SN 1997D and SN 1999br are listed in
Table~\ref{tab1}. Only the values of the radius $R_0$, mass
$M_{env}$, $^{56}$Ni mass $M_{Ni}$ and outer velocity $V_0$ (and
hence initial energy) of the envelope have been significantly varied.
The other model
parameters were maintained essentially equal to the values required to
reproduce the observations of SN 1987A. In particular, a colour
correction factor $f_c=T_c/T_{eff}=1.1-1.3$ has been adopted to
reproduce the observed continuum temperature for all the
supernovae. Small colour corrections are rather common in stellar and
supernova atmospheres and are induced by distortions of the continuum
caused by radiative transfer processes (e.g. scattering, wavelength
dependent opacities; see e.g. Eastman, Schmidt \& Kirshner
\cite{esk96}).

Interpolating simultaneously the light curve and the evolution of the
line velocity and continuum temperature results in a rather robust
``fit''. The estimated value of the initial thermal+kinetic energy of
the ejecta $E$ indicates that both events were rather under-energetic
compared to a typical Type II supernova. The inferred $^{56}$Ni mass
of SN 1999br is extremely small, testifying that the energy available
to produce and eject nucleosynthetic elements was very
low. Furthermore, because of the low luminosity in the plateau stage,
the post-explosion envelope is rather compact. The ejected envelope
masses are quite large, comparable to those required to reproduce the
plateau of typical Type II supernovae. In particular, the ejected envelope mass
of SN 1997D is almost three times larger than that estimated by Chugai
\& Utrobin \cite{cu00} and only 30\% smaller than that inferred by
Turatto et al. \cite{tetal98}. It is worth noting that the gross
properties of the light curve and line velocity of SN 1997D and SN
1999br can be roughly accounted for by SN 1987A-like parameters, but
simply decreasing the expansion velocity $V_0$ (and hence the energy
$E$) and $^{56}$Ni mass.

Although it is not straightforward to determine the error in the
parameters, we estimate that the intrinsic uncertainty of the ``fit''
is not larger than $\sim$ 30\% (see also Zampieri et
al. \cite{zetal02}). Additional sources of systematic errors are
related to the approximations introduced in the present analysis, in
particular to the choice of the initial conditions. Although it is
known that significant mixing occurred in SN 1987A (see e.g. Woosley
\cite{w88}), this effect may be less pronounced in other supernovae.
The light curve is very sensitive to the prescription for mixing and
to the actual velocity distribution as a function of mass $V(m)$. In
particular, the luminosity and duration of the plateau depend on the
energy and mass of the high velocity, Hydrogen rich part of the
envelope. If the innermost Helium and heavier elements layers did not mix
appreciably with the Hydrogen envelope, they would not produce any observable
effect (having very low energy). In this case, the estimated value of
$M_{env}$ would simply refer to the Hydrogen envelope mass. Then, in
general, $M_{env}$ represents a lower limit to the total mass of the
ejecta. Furthermore, the estimate of the envelope mass is sensitive to
uncertainties in the actual value of the gas opacity and the details
of the recombination physics. Here we adopted $\kappa = 0.2$ cm$^2$
g$^{-1}$, $t_{rec,0} \simeq 30$ days and $T_{eff} = 4000-4400$ K,
values similar to those used for the ``fit'' of SN 1987A. Only a
larger opacity and/or a delayed onset of recombination could in
principle allow for smaller envelope masses, but this is not in
agreement with what found for SN 1987A and other Type II supernovae
(Zampieri et al. \cite{zetal02}). The fact that the line velocities at
$\sim 100$ days are slightly larger than what predicted by the model
(see Figure~\ref{fig3}) seems to indicate that the assumption of
spatially constant density is only approximately correct and that more
mass is concentrated in the innermost, low velocity part of the
envelope. This effect may reduce the estimated energy $E$ because the
bulk of the kinetic energy is carried by the outer, high velocity
layers.

We emphasize that the present analysis for SN 1997D is based on the
assumption that it is an event similar to SN 1999br. However, although
the inferred parameters of the ejecta of SN 1997D rely on this
hypothesis, the estimates for SN 1999br are certainly valid because
they are not affected by uncertainties on the early light curve, the
duration of the plateau and the evolution of the line velocity.


Determining the mass of the progenitor is quite a difficult task and,
unless a pre-discovery identification is available (see, e.g., Smartt
et al. \cite{sma02}), the inferred value is usually rather uncertain.
Adding to the mass of the envelope reported in Table~\ref{tab1} the mass
of the collapsed core ($\approx 2 M_\odot$), the progenitors of SN
1997D and SN 1999br have at least 19 and 16 $M_\odot$,
respectively. We stress that these are most probably lower bounds and
the actual values of the progenitor masses are likely to be
larger. Therefore, the present estimates situate SN 1997D and SN
1999br in the intermediate mass progenitor range and rule then out an
origin from the low end of the mass range of core-collapse supernovae.

It should be noted that the results of the present analysis are
essentially insensitive to the inclusion of Rayleigh scattering from
neutral Hydrogen. In fact, towards the end of the plateau, $\tau_{ra}
\sim \kappa_{ra} \rho r \sim 0.1$, where $r\sim 10^{15}$ cm, $\rho
\sim 10^{-12}$ g cm$^{-3}$ and the Rayleigh scattering opacity
$\kappa_{ra}\sim 10^{-4}$ cm$^2$ g$^{-1}$. Therefore, the outer
recombined region is essentially transparent to Rayleigh scattering.

Understanding why the energy of these peculiar Type II supernovae is
low is of paramount importance in connection with the hydrodynamics of
the explosion and the formation of the central compact object. Because
the mass of the iron core before the explosion does not vary
significantly with progenitor mass $M_*$ (Woosley \& Weaver
\cite{ww95}), the gravitational potential energy liberated during the
collapse of the core is roughly independent of $M_*$. Therefore, the
energy of the ejecta depends mostly on neutrino reheating and the
hydrodynamic interaction of the supernova shock with the surrounding
material. To unbind a spherically symmetric envelope, the shock must
overcome the ram pressure of the gas that started to accrete after the
collapse of the core (see, e.g., Fryer \cite{fr99}). With increasing
progenitor mass both the ram pressure and the binding energy of the
envelope become larger.
Therefore, comparatively less energy remains available to heat up and
accelerate the ejecta. The final luminosity and expansion velocity are
then small and the resulting supernova is under-energetic (Zampieri
\cite{zamp02}). In this respect, it is tempting to note that the
difference between the inferred total (kinetic+thermal+binding) energy
of SN 1987A ($\sim 2.5 \times 10^{51}$ erg) and the ejected
kinetic+thermal energy of SN 1997D is $\sim 1.5 \times 10^{51}$ erg,
comparable to the binding energy of the envelope in a $25-30 M_\odot$
progenitor star (Woosley \& Weaver \cite{ww95}). Therefore, in SN
1997D, a large fraction of the gravitational potential energy
liberated by the collapse of the core may have been spent in trying to
unbind the massive envelope.

An important consequence of this scenario is that, because of the low
kinetic energy acquired by the envelope, a large fraction of the
stellar material (in particular the innermost layers that have smaller
velocities) is likely to remain gravitationally bound to the core
after shock passage and fall back onto it. This is confirmed also by
the small amount of $^{56}$Co present in the ejecta. The fallback of
stellar material may also turn the newly formed neutron star into a
black hole. If this happens, the late time fallback onto the central
black hole may give rise to detectable emission of radiation with a
characteristic power-law decay (Zampieri et al. \cite{zampetal98},
Balberg et al. \cite{balbetal00}). The detection of this emission in
the late time light curve would provide the first direct evidence for
the presence of a black hole in the site of its formation.

We note that the overall picture may be different for a supernova from
a massive progenitor whose post-shock envelope retains a significant
fraction of its initial angular momentum. In this case, the
hydrodynamics of the explosion becomes more complex and may give rise
to a high energy, asymmetric explosion (MacFadyen \& Woosley
\cite{macw99}).



The present analysis indicates that the parameters of the ejecta of SN
1997D and SN 1999br are consistent with them being intermediate mass
core-collapse supernovae. If the mass of their progenitors is
sufficiently large, they could form a black hole remnant, giving rise
to significant fallback and late-time accretion onto the central
compact object. In order to confirm these findings, more detailed
investigations using radiation hydrodynamic simulations and spectral
synthesis calculations are being planned. Clearly, monitoring this
type of supernovae from discovery to late phases is of the utmost
importance.

\bigskip

\noindent
{\bf ACKNOWLEDGMENTS}
\smallskip

We thank the referee Stan Woosley for valuable comments. We
acknowledge support from the Italian Ministry for Instruction,
University and Research (MIUR) through grant Cofin MM02905817 and the
Italian Space Agency (ASI) under grant ASI
I/R/70/00. M. H. acknowledges support for this work by NASA through
Hubble Fellowship grant HST-HF-01139.01-A awarded by the Space
Telescope Science Institute, which is operated by the Association of
Universities for Research in Astronomy, Inc., for NASA, under contract
NAS 5-26555.

\label{lastpage}

\end{document}